\documentstyle[psfig,mncite]{mn}

\begin{document}
\title[Can Jupiters be found by monitoring Galactic Bulge microlensing events from northern sites?]{Can
Jupiters be found by monitoring Galactic Bulge microlensing events from northern sites?
\thanks{Based on observations made with the IAC 0.8m telescope at Izana
Observatory, Tenerife, operated by the Instituto de
Astrofisica de Canarias.}}

\author[Tsapras, Street et al.]
{\parbox[t]{\textwidth}{Yiannis~Tsapras$^1$, Rachel A.~Street$^1$, Keith~Horne$^1$ $^2$, Alan~Penny$^3$,
Fraser~Clarke$^4$, Hans Deeg$^5$ $^6$, Francisco Garzon$^5$, Simon Kemp$^5$, Maria Rosa Zapatero
Osorio$^5$, Alejandro Oscoz Abad$^5$, Santiago Madruga Sanchez$^5$, Carlos Eiroa $^7$, Alcione Mora $7$,
Antxon Alberdi $^6$, Andrew Cameron $^1$, John K. Davies$^8$, Roger Ferlet $^9$, Carol Grady $^1$$^0$,
Allan W. Harris $^1$$^1$, Javier Palacios $^7$, Andreas Quirrenbach $^1$$^2$, Heike Rauer $^1$$^1$, Jean
Schneider $^1$$^3$, Dolf de Winter $^1$$^4$, Bruno Merin $^1$$^5$, Enrique Solano $^1$$^5$} \\
$^1$School of Physics and Astronomy, Univ. of St Andrews, Scotland KY16 9SS \\
$^2$Department of Astronomy, University of Texas, Austin TX 78712, USA\\
$^3$Rutherford Appleton Laboratories, Oxon, England \\
$^4$Institute of Astronomy, Cambridge, England \\
$^5$Instituto de Astrofisica de Canarias, Canary Islands, Spain \\
$^6$Instituto de Astrofisica de Andalucia, Granada, Spain \\
$^7$Dpto. Fisica Teorica. Universidad Autonoma de Madrid. Spain\\
$^8$Joint Astronomy Center, Hilo, Hawaii, USA \\
$^9$Institute d'Astrophysique, CNRS, Paris, France\\
$^1$$^0$Eureka Scientific, USA\\
$^1$$^1$Institute of Planetary Exploration, DLR, Berlin, Germany\\
$^1$$^2$University of California, San Diego, La Jolla, CA, USA\\
$^1$$^3$Observatoire de Paris, Meudon, France \\
$^1$$^4$Space Research Organization of the Netherlands, Groningen, The Netherlands\\
$^1$$^5$Laboratorio Astrofisica Espacial y Fisica Fundamental, Madrid, Spain \\
}

\date{submitted May 2000}
\maketitle

\begin{abstract}
In 1998 the EXPORT team monitored microlensing event lightcurves using a CCD camera
on the IAC 0.8m telescope on Tenerife to evaluate the prospect of using northern telescopes
to find microlens anomalies that reveal planets orbiting the lens stars. The high airmass and more limited
time available for observations of Galactic Bulge sources makes a northern site less favourable for
microlensing planet searches. However, there are potentially a large number of northern 1m class telescopes
that could devote a few hours per night to monitor ongoing microlensing events. 
Our IAC observations indicate that accuracies sufficient to detect planets can be achieved despite the
higher airmass. 
\end{abstract}


\section{INTRODUCTION}

In 1995, Mayor and Queloz reported the detection of a planet orbiting the star 51
Peg.  This was the first report of a planetary companion to a normal star outside
the solar system, and was quickly followed by other discoveries
\cite{marcy96}.  Even prior to that, \scite{wolsz} reported the discovery
of three planet-mass objects orbiting the pulsar PSR1257+12, revealing their
presence through perdiodic variations in the arrival times of radio pulses from the 
star. Since then, reports of new objects orbiting distant stars have been steadily
increasing ({\it http://www.obspm.fr/encycl/encycl.html}). 

In these last few years, several search groups have been formed utilising a
variety of observing techniques to increase the number of detections and
place meaningful statistics on the type and number of planets orbiting normal
stars. One such technique is microlensing \cite{pacz96,albrow98}, which 
probes the `lensing zone', $\sim1-4$~AU for a typical $0.3~M_\odot$ lens star.
Microlensing is unique among ground-based techniques in its sensitivity to 
low-mass planets down to the mass of Earth
\cite{bennet96}.

\subsection{MICROLENSING BASICS}

Microlensing involves the gravitational deflection of light from a background
star (source) as a massive stellar object (lens) passes in front of it. This
results in two images of the background source, on opposite sides of the lens
position. For sources in the Galactic Bulge, the image separation is $\sim10^{-3}$~arcsec 
and thus unresolvable. 
What is actually observed in microlensing events is a variation of
the brightness of the source star as the lens moves in front of  it. Since more
light is bent towards the observer, the combined brightness of the two lensed
images is greater than that of the unlensed source.  The total amplification is
given by:
\begin{equation}
A = \frac {u^{2} + 2} {u(u^{2} + 4)^{1/2}},
\end{equation}
where ${u = R_{S}/R_{E} = \left( u_{\rm min}^2 + \left( \frac{2(t - t_{0})}{t_E}\right) \right)^{1/2}}$, ${R_{S}}$ is the separation on the lens plane between the 
source and the lens, and ${R_{E}}$ is the Einstein ring radius of the lens, given by
\begin{equation}
R_{E} = \sqrt{\frac{4GM_{L}D_{L}D_{LS}} {c^{2}D_{S}}}.
\end{equation}
${D_{LS}, D_{S}, D_{L}}$ are the lens-source, observer-source and observer-lens
distances respectively \cite{pacz86}. Also, $t_{0}$ is the time of maximum amplification 
and ${t_{E}}$ the event timescale.

Galactic Bulge lensing events have typical timescales ${t_{E}=\frac{2R_{E}}{v_{\perp}}=}$ 
10-100 days, where ${v_{\perp} \sim 200}$ km~s$^{-1}$ is the transverse velocity between 
the source and lens and ${t_{E}}$ is the time to cross the diameter of the Einstein 
ring \cite{bennet96}. If a planet orbits the lens star within the
`lensing zone', ${0.6 \le a/R_{E} \le 1.6}$ ({\it a}
being the transverse component of the planetary orbital radius), then binary
lensing may produce a light-curve that deviates by a detectable amount from the single-lens case
\cite{gould92}. By correctly assessing such light-curve deviations (or
anomalies), the presence of planetary bodies can be deduced \cite{bennet96,pacz96}.

The Einstein ring radius for a solar mass lens
half-way to the galactic centre is about 4 AU. This is close to the orbital radius of Jupiter
from the Sun. The event duration scales with the size of the Einstein ring, and hence as ${\sqrt{m_{p}}}$.
Lensing by a Jupiter-mass planet with $q = m_p/M_L \sim 3\times 10^{-3}$ will therefore be some 20 times
briefer than the associated stellar lensing event, hence typically 0.5 - 5 days.

We can crudely estimate the planet detection probability assuming that the planet
is detected when one of the two images of the source falls inside the planet's
Einstein ring. This turns out to be ${\sim 20\%}$ for a Jupiter and ${\sim 2\%}$ for Earths. 

The fitting of theoretical models to the lightcurve yields the mass ratio and
normalised projected orbital radius for the binary lens \cite{gould92}. A number
of collaborations have formed to perform yearly systematic searches for
microlensing events, by repeatedly imaging starfields towards the Galactic
Centre \cite{MACHO,OGLE}.  This offers both rich background starfields
and lensing objects at intermediate distances. Microlensing events are being
reported regularly via internet alerts issued by a number of collaborations (MACHO - now terminated,
OGLE, EROS).

\section{A STRATEGY FOR FINDING JUPITERS}
To discover and quantify planetary anomalies in a light curve, events in progress
must be imaged very frequently. To
correctly estimate the duration and structure of the anomalous peak, and thus
measure the planetary mass and position relative to the lens, we
require many photometric measurements during the anomalous deviation. Ideally, a search 
for Jupiters would employ hourly imaging, 
which also increases the possibility of detecting deviations caused by Earth-mass 
planetary companions, whose deviations last only for a few hours. However, daily 
sampling from a northern site might already suffice to {\it detect} Jupiters, if not to
characterize them.

In 1998, over one hundred alerts were issued by the MACHO and OGLE teams. 
Let us assume that 15\% of solar type stars 
have Jupiters within the lensing zone. Only 20\% of those will produce detectable 
deviations \cite{gould92}, since most of the time the planet will not be near the image trajectories. 
We then expect that ${\sim 3}$ of the 100 events reported in 1998 had Jupiter deviations. 
The question that arises is whether and how accurately would we be able to detect them 
with observations from northern sites ? 

Let us adopt the aforementioned assumption and assume additionally that we have access to a
1m class telescope at ${+30^{\circ} }$ latitude. Then we have a 3 hour observing window for the Bulge for 
a period of 4 months. If the mean exposure time is 600~s and the CCD readout time is 180~s, 
then we should be able to make 14 exposures per night, and thus follow a maximum of 14 events with
one image per night. Since on the important events we would require more than 1 data point per night we 
can cut the number of events followed down to 9 events per night.

Observations should intensify, by re-allocating the nightly imaging of different targets, at 
times around the time of maximum amplification and events
should be followed in order of importance, i.e. an event is given higher
priority if it is close to maximum amplification.

There were over 100 alerts issued in 1998, so the average number of microlesing events 
in the 4 months that the Bulge could be observed from the North would be ${\sim 35}$.
If each event was imaged for ${\sim 30}$ days then
these events could have been covered intensively enough to detect any giant 
planet deviations that might have occured close to the time of maximum amplification when 
such deviations are more pronounced. 

Deviations due to giant planets last for a few days \cite{gould92}, so with daily
monitoring we should get one or two data points deviating from the unperturbed
light-curve. Therefore if any of the 35 events observed had a giant planet in the
lensing zone (under our previous assumption, one event should) it ought to be detectable. Furthermore, if a series 
of telescopes were dedicated to this task in coordinated
operation, the temporal coverage of the events and/or the number of events
observed would be increased. 
 
 If daily sampling suffices to detect most of the short lived lensing anomalies due to
 Jupiters, more intensive monitoring is necessary if the planetary characteristics are
 also to be determined. 
 The planet/star mass ratio is the square of the event
 durations and the shape of the anomaly identifies which image of the star is being lensed
 by the planet. Characterization requires perhaps 5-10 points/night spanning the duration of the 
 anomaly. For this reason
 current lensing searches with Southern telescopes have aimed for hourly sampling of the
 most favourable events. Prompt automatic data reduction and  internet alerts would be an
 alternative method of triggering continuous monitoring within minutes after an 
 anomaly is found. This 2-level strategy would allow more events to be monitored for
 Jupiters.

\section{OBSERVATIONS SUMMARY}
It remains to be demonstrated whether useful photometric measurements 
can be achieved at northern sites. At ${+30^{\circ} }$ latitude, airmass is below
2 for only 3 hours per night. As atmospheric transmission and seeing are poorer at large airmasses, 
it is not obvious that sufficient accuracy to characterize the microlensing lightcurves for 
Galactic Bulge sources can be achieved from a northern site.

We gathered data in 1998 looking at microlensing events in the
Galactic Bulge.  The {\em IAC} 0.8m telescope on Tenerife (Longitude:
${16^{\circ}30'35"}$ West, Latitude: ${28^{\circ}18'00"}$ North) in the Canary Islands was used for one 
hour per night for a period of 4.5 months (15 May-30 Sept). Several ongoing
microlensing events were monitored with 1 or 2 being observed each night. 

\begin{figure}
\centering
\begin{tabular}{c}
\psfig{file=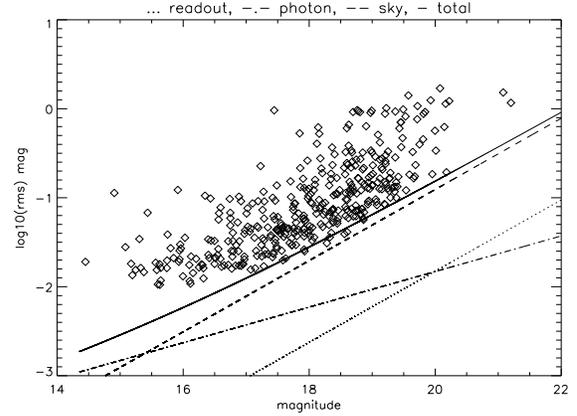,angle=0,width=8cm}
\end{tabular}
\caption{\small Magnitude values versus the corresponding rms values of 15
measurements of the magnitude values for 390 stars. The plot looks more noisy than expected which is due to the overcrowding of some stars.}
\protect\label{sigscat}
\end{figure}

In the observing run, the number of nights per event ranged from 3
to 15, with a maximum of 3 images per night taken at 10~min
intervals. Exposure times were 600~s for each image and
all were obtained in the R-band. The CCD size was 1024 ${\times}$ 1024, covering a sky area of 7.3 ${\times}$ 7.3 arcminutes and the typical seeing ranged between 1.5 and 2 arcsec. The microlensing events were recorded with a
 photometric accuracy that reached ${\sim1-2}$\% (see Fig.~\ref{sigscat}) for
 the brighter part of the light-curve (${R\sim16}$~mag) but no planetary deviations from the event light curves were found. This was not unexpected since the gaps in
temporal sampling were of appreciable size. The two best sampled events are discussed
in section 5.

\section{CROWDED FIELD PHOTOMETRY}

We performed crowded field photometry on the CCD data using the {\sc starman} stellar
photometry package \cite{starman} in a semi-automated data
reduction pipeline. Further processing of these results and lightcurve analysis
was performed by means of programs developed by the authors.

The CCD frames were de-biased and flat-fielded and the target was identified
from finder charts. A coordinate list of stars selected for photometry was
compiled manually.  This list included the target star, ${\sim 20}$ bright,
unsaturated stars which were used to calibrate the point spread function (henceforth called the
PSF stars) and a selection of stars of constant brightness comparable to that of
the target at each stage of the lensing event (henceforth called the error
stars). The latter were used to calculate the RMS scatter on the measured
target magnitude for the full range of its brightness variation. The list also included 
any close companions to the aforementioned stars, which  might otherwise distort the PSF 
fitting photometry if ignored.

\begin{figure}
\centering
\begin{tabular}{c}
\psfig{file=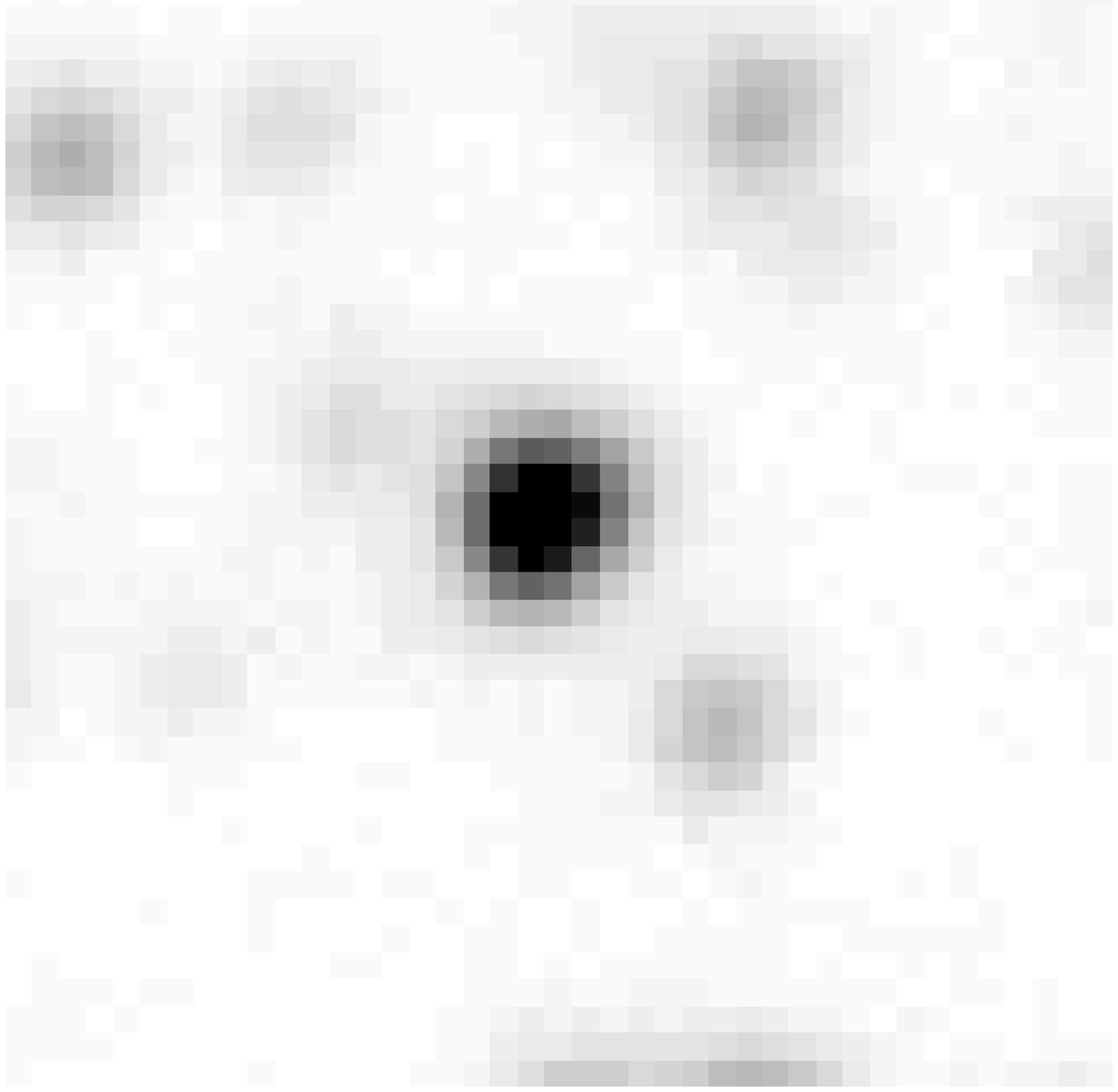,angle=0,width=2cm,height=2cm}
\psfig{file=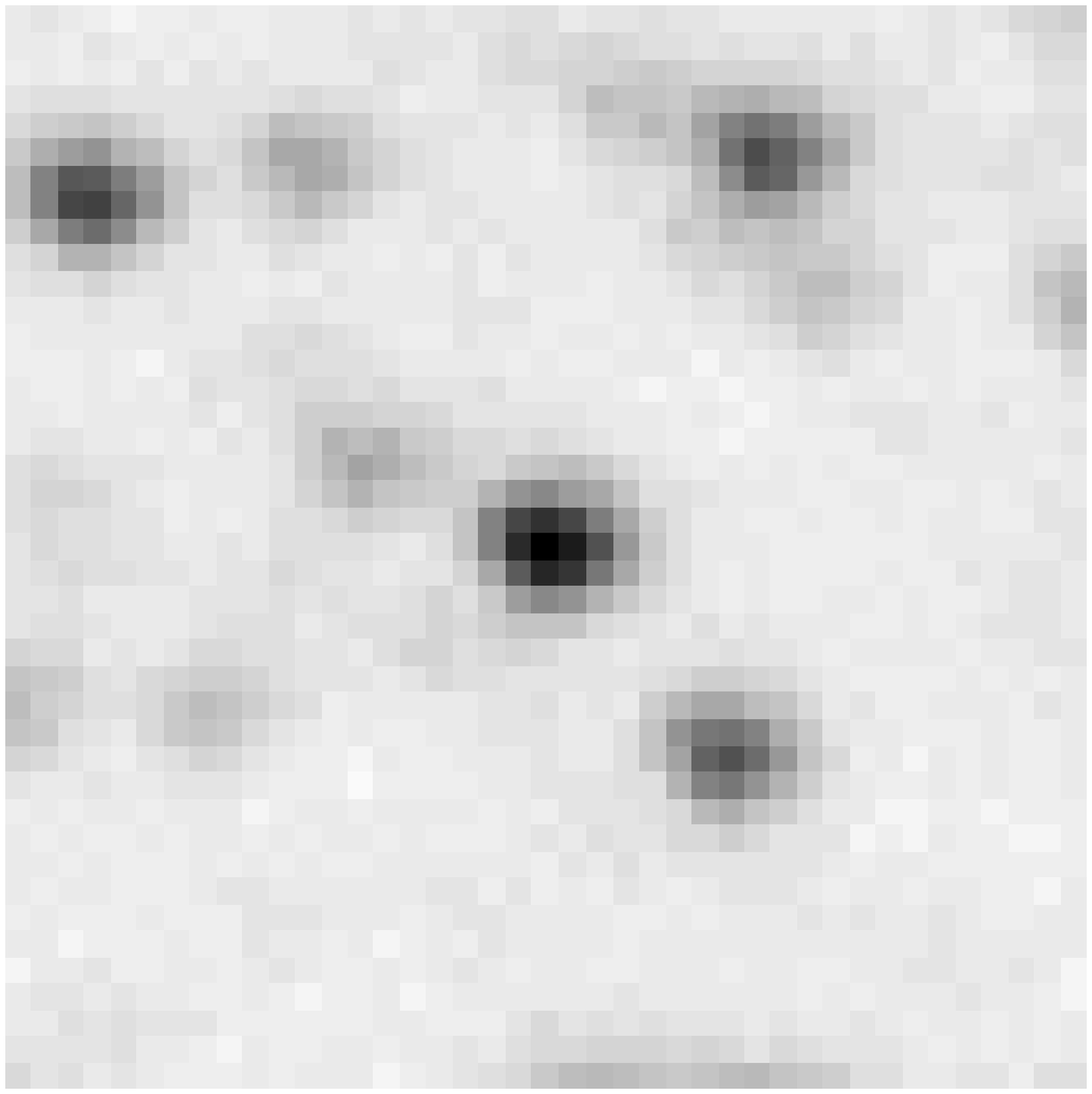,angle=0,width=2cm,height=2cm}
\psfig{file=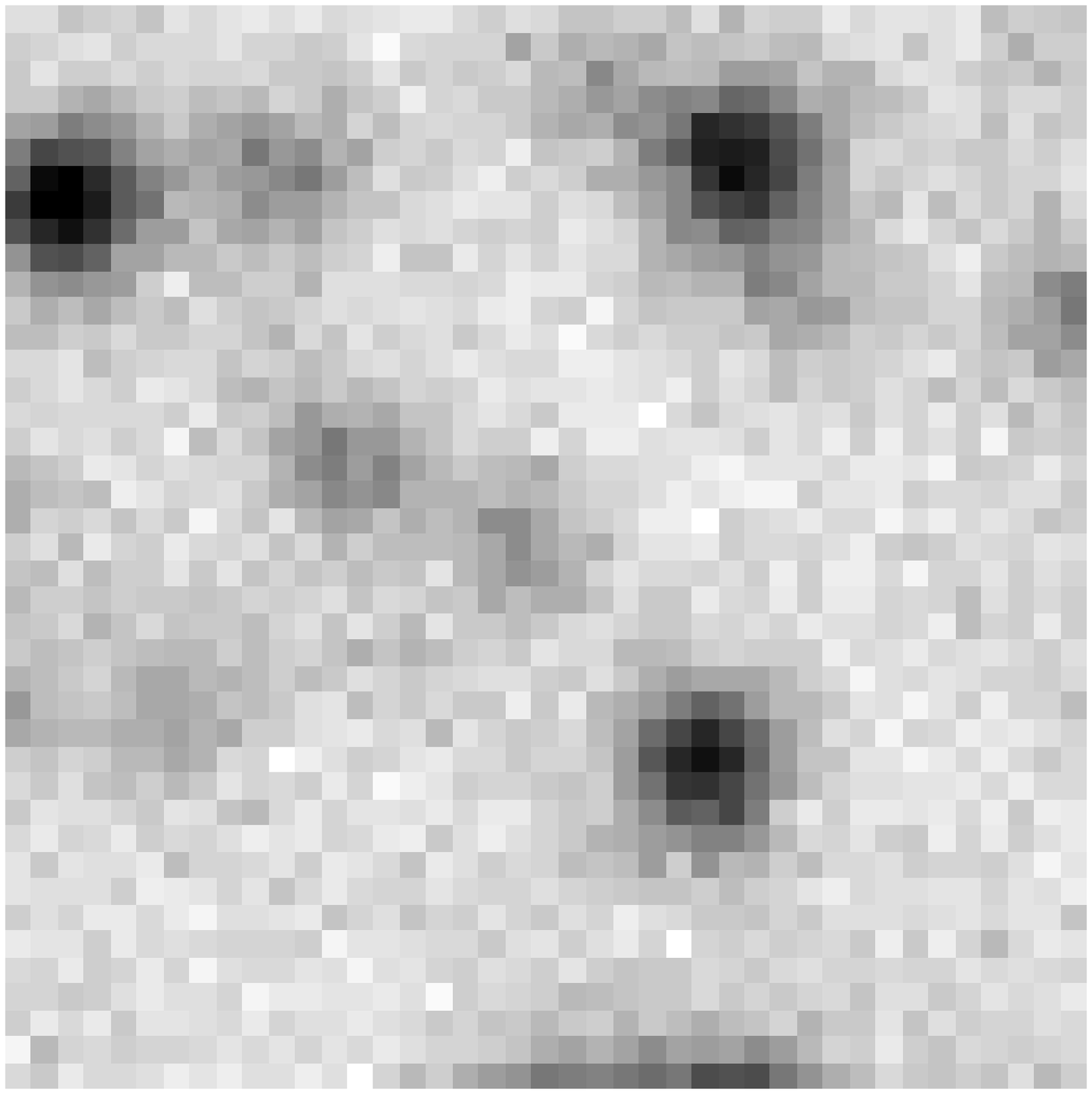,angle=0,width=2cm,height=2cm}
\psfig{file=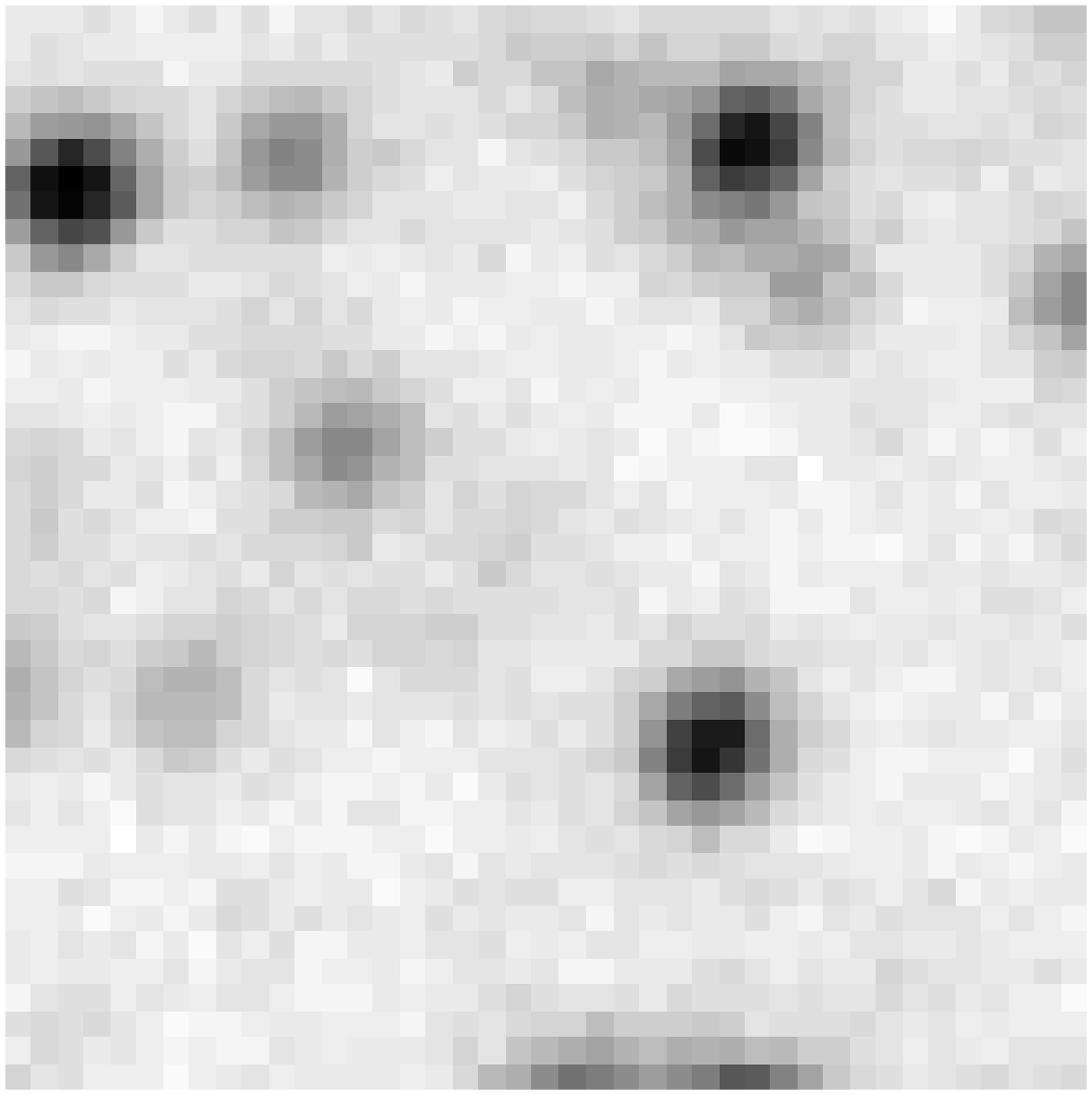,angle=0,width=2cm,height=2cm}
\end{tabular}
\caption{\small Progress of microlensing event 98BLG35. The box sizes are 40
${\times}$ 40 pixels. North is right, East is down.}
\protect\label{days}
\end{figure}

The images were registered using {\sc figaro} to determine relative pixel shifts in 
the ${x}$ and ${y}$ axes for each frame.  Automated cropping was performed on each image,
creating a sub-frame, such that the star list was correctly aligned for each
sub-frame. A PSF profile was then derived from fitting to the the PSF stars. 

Crowded field PSF-fitting photometry was performed on the stars in the main list. 
Stars with poor PSF fits were rejected. The magnitudes of the PSF stars were measured 
separately. These stars were used to set the zero point of the instrumental magnitudes, 
since these bright, isolated stars are less affected by photon noise or close companions. 
Differential magnitudes for the stars in each field were measured relative to the average flux of the 
PSF stars.
Although no standard stars were observed, we have added a constant to the
{\sc starman} instrumental magnitudes to make them match the baseline $R$ magnitudes reported by the MACHO team
({\it http://darkstar.astro.washington.edu/}) to an accuracy of 0.1~mag.

To quantify the accuracy of our differential photometry we calculated for 390 stars in the field of
98BLG35 the RMS scatter about the weighted mean of 15 measured magnitudes. Fig.~\ref{sigscat} 
shows the resulting estimate of the rms magnitude error as a function of the star's $R$ magnitude.
The vertical scatter of the points at a given $R$ magnitude in Fig.~\ref{sigscat} is consistent with
the uncertainty ($\sqrt{2/(N-1)} \sim 0.4$) given that our estimate of the rms magnitude error is based on
$N=15$ measurements of each star. The achieved accuracy is some 3 times worse than expected
based on our CCD noise model (curves in Fig.~\ref{sigscat}), which is dominated by sky noise for stars fainter
than $R \sim 16$~mag. We attribute the degradation of accuracy to the effects of crowding, where the PSF-fit
has difficulty separating contributions from blended star images.

Fig.~\ref{sigscat} indicates that our 600~s exposures have achieved an accuracy approaching 1-3\% for
well-exposed images of brighter $R \sim 16$~mag stars. The achieved photometry degrades to 10\% at $R \sim
18-19$. This accuracy can theoretically be improved by applying a seeing correction to the data sets.
However, we found no obvious correlation of magnitude residuals with seeing or sky brightness.
It is probably possible to further improve the accuracy of our differential photometry by further refinement
of the analysis techniques, for example by means of image subtraction methods \cite{alard98} which have recently
been demonstrated to get close to theoretical limits. However, the accuracy we have achieved is already 
sufficient for detection of planetary lensing anomalies, as we now demonstrate.

\section{RESULTS FOR 98BLG35 AND 98BLG42}

\begin{figure}
\centering
\begin{tabular}{c}
\psfig{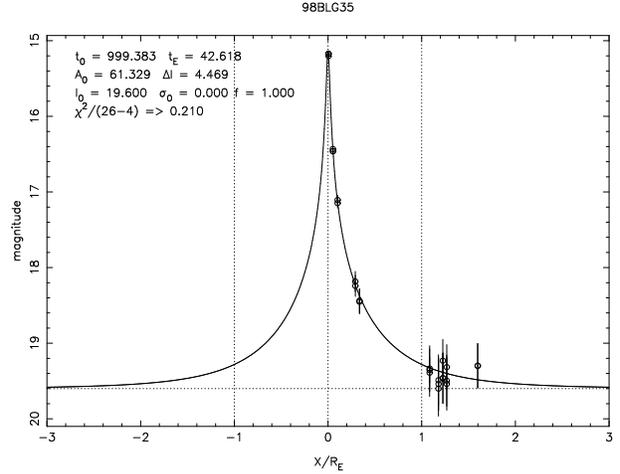}
\end{tabular}
\caption{\small Fitted lightcurve for microlensing event 98BLG35. 
R-Magnitude is plotted versus separation in units of ${R_{E}}$. 
The estimated event parameters are shown on the top left corner of the plot.
}
\protect\label{98BLG35}
\end{figure}

\begin{figure}
\centering
\begin{tabular}{c}
\psfig{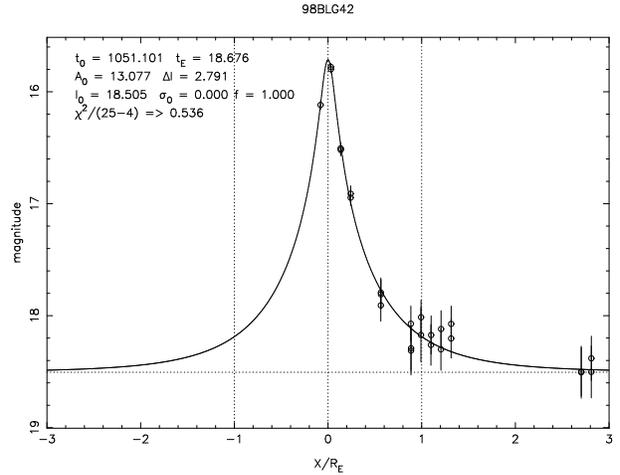}
\end{tabular}
\caption{\small Fitted lightcurve for microlensing event 98BLG42. R-Magnitude
is plotted versus separation in units of ${R_{E}}$. The slight increase in brightness in the region
${x/R_{E} \sim 1}$ of the plot is probably a blending effect from a star that lies almost on top 
of the target. As the target gets very faint the PSF-fitting program has
difficulty distinguishing between the two stars.}
\protect\label{98BLG42}
\end{figure}

Our light curves for MACHO 98BLG35 (Fig.~\ref{days} presents four frames showing the
progress of the event) and 98BLG42 were the best-sampled events and will be 
discussed here. The observations for these events started near
maximum amplification (see Fig.~\ref{98BLG35} and Fig.~\ref{98BLG42} with
estimated event parameters: 
time of maximum amplification, event timescale, maximum amplification and
baseline magnitude ${t_{0} ,t_{E} , A_{0}, I_{0}}$  respectively at the top left of the plot).
The photometric analysis details are presented in Table~\ref{tab:98blg} for 
both events.

\begin{table*}
\centering
\caption{98BLG35 \& 98BLG42 Observations}
\protect\label{tab:98blg}
\vspace{5mm}
\begin{tabular}{lccccl}
\hline
	HJD (245+)  & R Magnitude  98BLG35 & Magnitude error  & HJD (245+)  & R Magnitude 98BLG42 & Magnitude error\\
\hline
       0999.486  &    15.176	   &	    0.015  &	      1050.356  &    16.117	 &	    0.039  \\
       0999.493  &    15.169	   &	    0.015  &	      1051.380  &    15.777	 &	    0.031  \\
       0999.501  &    15.202	   &	    0.016  &	      1051.387  &    15.799	 &	    0.032  \\
       1000.498  &    16.432	   &	    0.038  &	      1052.359  &    16.505	 &	    0.052  \\
       1000.506  &    16.461	   &	    0.039  &	      1052.369  &    16.519	 &	    0.052  \\
       1000.513  &    16.460	   &	    0.039  &	      1053.355  &    16.945	 &	    0.071  \\
       1001.559  &    17.147	   &	    0.064  &	      1053.363  &    16.910	 &	    0.069  \\
       1001.567  &    17.106	   &	    0.062  &	      1056.360  &    17.907	 &	    0.141  \\
       1005.585  &    18.183	   &	    0.133  &	      1056.368  &    17.793	 &	    0.130  \\
       1005.593  &    18.241	   &	    0.139  &	      1056.375  &    17.809	 &	    0.132  \\
       1006.541  &    18.451	   &	    0.161  &	      1059.361  &    18.072	 &	    0.158  \\
       1006.549  &    18.438	   &	    0.159  &	      1059.368  &    18.312	 &	    0.217  \\
       1022.486  &    19.354	   &	    0.165  &	      1059.376  &    18.291	 &	    0.199  \\
       1022.493  &    19.389	   &	    0.313  &	      1060.365  &    18.171	 &	    0.243  \\
       1022.501  &    19.335	   &	    0.301  &	      1060.372  &    18.014	 &	    0.152  \\
       1024.506  &    19.600	   &	    0.364  &	      1061.356  &    18.261	 &	    0.181  \\
       1024.513  &    19.490	   &	    0.336  &	      1061.375  &    18.172	 &	    0.170  \\
       1024.521  &    19.540	   &	    0.348  &	      1062.357  &    18.300	 &	    0.187  \\
       1025.497  &    19.230	   &	    0.280  &	      1062.364  &    18.119	 &	    0.164  \\
       1025.504  &    19.463	   &	    0.330  &	      1063.367  &    18.203	 &	    0.174  \\
       1025.512  &    19.463	   &	    0.336  &	      1063.375  &    18.073	 &	    0.159  \\
       1026.383  &    19.495	   &	    0.338  &	      1076.340  &    18.505	 &	    0.232  \\
       1026.391  &    19.538	   &	    0.348  &	      1076.347  &    18.501	 &	    0.215  \\
       1026.398  &    19.316	   &	    0.297  &	      1077.339  &    18.502	 &	    0.231  \\
       1033.419  &    19.297	   &	    0.293  &	      1077.347  &    18.381	 &	    0.198  \\
       1033.426  &    19.299	   &	    0.294  &   &           & \\

\hline
\end{tabular}
\end{table*}

A 2-10 Earth-mass planetary companion to the lensing star in 98BLG35 was suggested by the 
MPS/MOA team \cite{rhie98}. We are unable to confirm this since our lightcurve for this event 
covered only the decline and as a consequence
the peak was not clearly defined in the fit. Unfortunately all of the events
observed suffered from this same problem, with the exception of 98BLG42 where we
had one point before peak magnification. For this reason our fits to the data do not 
yield definite event parameters, but are nevertheless in agreement with the
ones reported by other follow-up teams that use a number of dedicated telescopes for
the same purpose. 

The PLANET group issued an anomaly alert for 98BLG42 claiming it to be the result of 
binary lensing with finite source effects. They report an anomalous decline that 
occured between JD 2451050.5 and 2451051.2, close to the time of maximum 
amplification, attributable to a caustic crossing by a resolved source. 
We have obtained 2 observations at JD 2451051.3804 and JD 2451051.3879
but are unable to confirm anything since we do not detect any
significant deviations from the unperturbed lightcurve. As far as we are aware, no data have as yet been
published for this event.

\begin{figure}
\centering
\begin{tabular}{c}
\psfig{file=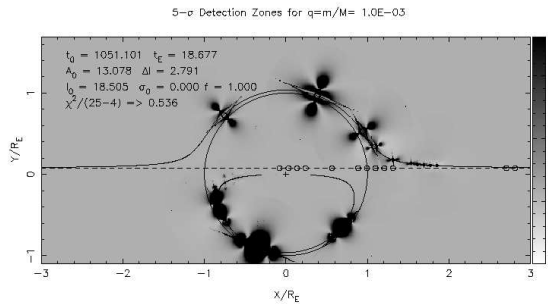,angle=-90,width=8cm}
\end{tabular}
\caption{\small ${\Delta\chi^{2}}$ -vs- planet position for the data on 98BLG42. The black zones 
show where the presence of a planet with $q = 10^{-3}$ is ruled out by our observations.}
\protect\label{chi42}
\end{figure}

Fig.~\ref{chi42} shows a ${\Delta\chi^{2}}$ map as a function of planet position with $q = 10^{-3}$ for the event 98BLG42.
Our first 4 observations of this event occur at 1 day intervals, followed by two 3-day gaps between
the next 2 data points. This is a relatively high amplification event and therefore the images of the
source star move quite rapidly around the Einstein ring. For this reason the `detection zones' set by
our observations at 1-day intervals do not overlap. Although incomplete, we nevertheless do achieve a
significant detection probability.

The probability of finding a planet on position $x,y$ on the lens plane given its orbital radius 
$a$ (assuming a randomly oriented circular orbit) is given by:
\begin{equation}
P(det|a) = \int P(det|x,y) P(x,y|a) dx dy.
\end{equation} 
The first term
\begin{equation}
P(det|x,y) = 1 - \exp{\left(\frac{- \Delta \chi^{2} (x,y)}{2}\right)}
\end{equation}
is 0 in the `grey zones' on Fig.~\ref{chi42}, where a planet has no
effect on the lightcurve,
and 1 in the `black zones', where the planet produces
a large effect near one of the data points.
This detection probability is appreciable only when the
planet position $x,y$ is close to one of the images of the source
at the time of one of the data points in the lightcurve.
The interesting shape of the black zones in which the planet can
be detected is due to details of lensing by two point masses,
which we have calculated using the techniques of
Gould and Loeb (1992).

The second term $P(x,y|a)$ is obtained by randomly orienting the
planet's assumed circular orbit of radius $a$,
and then projecting it onto the $x,y$ plane of the sky.
This gives a circular distribution centred on the lens star
and rising as $(r/a)^2$ to a sharp peak at $r=a$, outside
which the probability vanishes. This term may be written as:
\begin{equation}
P(x,y|a) = \frac{1}{2 \pi a \sqrt{a^{2}-x^{2}-y^{2}}}
\end{equation}
for ${r^{2} = x^{2} + y^{2} < a^{2}}$.
A slightly elliptical orbit would blur out the outer edge, and it's
obviously possible to calculate this for any assumption about the
eccentricity.

\begin{figure}
\centering
\begin{tabular}{c}
\psfig{file=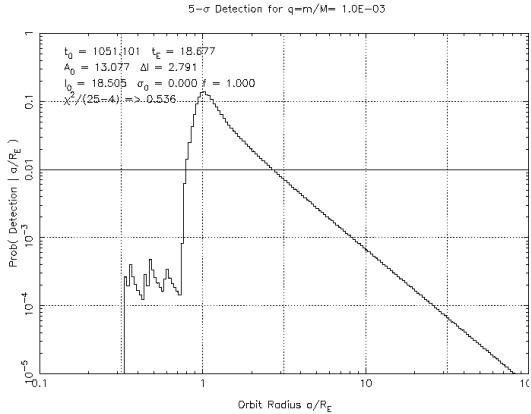,angle=-90,width=8cm}
\end{tabular}
\caption{\small The probability of detecting a planet with mass ratio
q=$10^{-3}$
in orbit at radius $\alpha$ in units of the Einstein ring radius based on the
observations for 98BLG42. The solid horizontal line indicates the total detection probability. The
probability is maximized for orbital radius ${a/R_{E} \sim 1}$. }
\protect\label{probdet}
\end{figure}

The net detection probability $P(det|a)$ is therefore the result
of summing up the fraction of the time that a planet in the
orbit of radius $a$ would be located inside one of the `black zones'
of Fig.~\ref{chi42}. The result is plotted in Fig.~\ref{probdet}.
Since the detection zones are near the lens star's Einstein ring,
the detection probability is highest for planets with $a\sim R_E$.

Our observations, primarily the data points on 4 consecutive
nights while the source was strongly amplified,
yield a detection probability of about 10\% for $a = R_E$.
This detection probability is for a planet with a Jupiter-like
mass ratio, $q=m_p/M_L=10^{-3}$, and for
other planet masses it scales roughly as $\sqrt{q}$.
For $a<R_E$ the detection probability in Fig.~\ref{probdet} is lower because
the planet spends more of its time inside the detection zones.
Discrete steps occur as the orbit radius shrinks inside each of
the data points.
For $a>R_E$ the planet spends most of its time outside the
detection zones and the probability drops off as $(R_E/a)^2$.

To summarize, our measurements of the lightcurve of 98BLG42
probe a substantial fraction of the lensing zone for the
presence of Jupiters.
Our detection probability, arising mainly from data on 4 consecutive
nights of high amplification,
is 10\% for a planet with a Jupiter-like mass ratio
$q=10^{-3}$ and orbit radius $a \sim R_{E}$. 
The gaps between our detection zones indicate that
denser temporal coverage would improve the result for this event
by perhaps a factor of 3.
For even denser sampling, however, the detection zones in Fig.~\ref{chi42}
would begin to overlap, diminishing the added value of each new data
point toward the objective of detecting Jupiters.

\section{SIMULATED DETECTION OF A JUPITER}

In this section we show explicitly how Jupiters can be detected in lightcurve data
obtainable from a northern site. Our goal is not to characterize the planet, but rather
to show that we can discover that a planet deviation has occured, based on the daily
sampling and accuracy attainable from a northern site.

To make a reasonably realistic assumption of our ability to detect planets,
 we add several fake data points to our observed lightcurve of 98BLG42. 
 These points fill in a 4-day gap in the actual observations
during the decline from peak amplification. The fake data points include the effect of a
 Jupiter mass planet located at ${x/R_{E}=1.05}$, ${y/R_{E}=0.39}$, which
amplifies the major image on one night only. The magnitudes reported in this section are {\sc starman} 
instrumental magnitudes.

The fake data points were obtained by using the lightcurve magnitude value for that 
day with an added random scatter value (${\Delta magnitude}$) within the limits 
imposed by the noise model.

\begin{figure}
\centering
\begin{tabular}{c}
\psfig{file=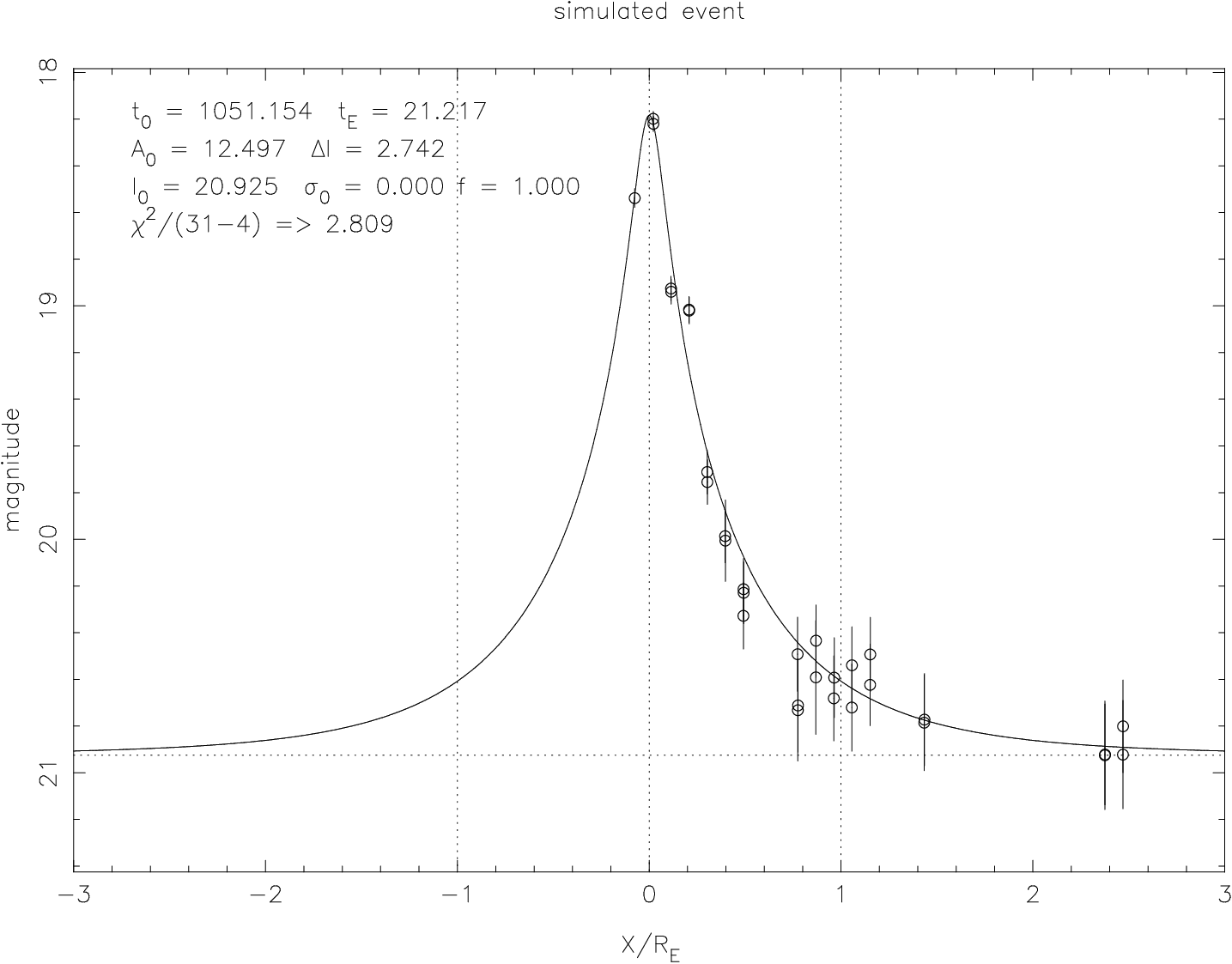,angle=-90,width=8cm}
\end{tabular}
\caption{\small Shown above is the best fit single lens model for a simulated lightcurve 
which includes a planetary deviation. The fitted parameters appear in the top left
corner of the plot. The residuals of the fit are shown in fig~\ref{resnoplan}.
The ${\chi^{2}}$ value improves by a factor of 8 if we allow for the presence of a 
planet as shown in fig~\ref{planet42}.}
\protect\label{single}
\centering
\begin{tabular}{c}
\psfig{file=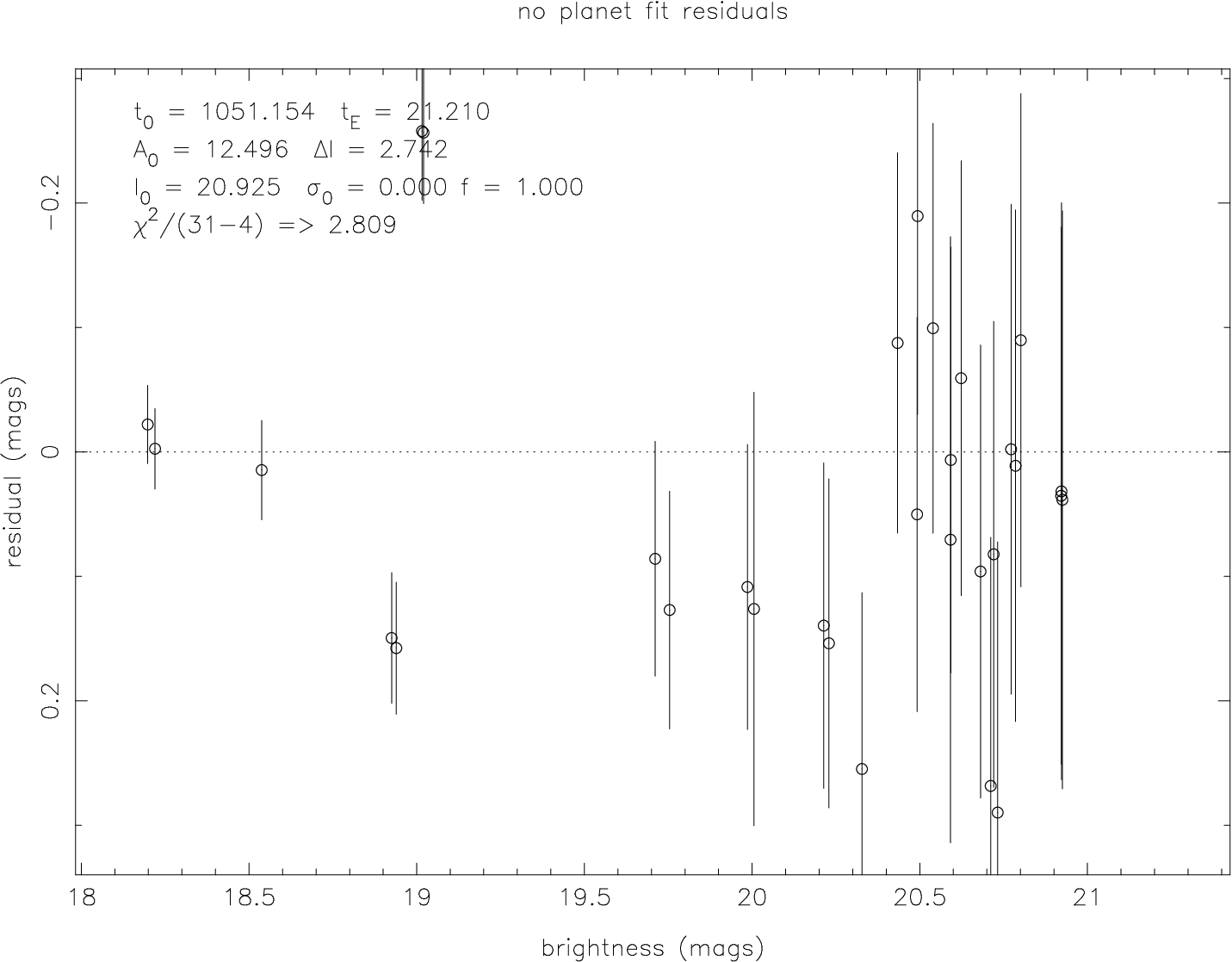,angle=-90,width=8cm}
\end{tabular}
\caption{\small Fit residuals for the best single lens model fit as shown is
fig~\ref{single}.}
\protect\label{resnoplan}
\end{figure}

The new lightcurve, including the fake data points and the best-fit point-lens
lightcurve, are shown in Fig.~\ref{single}. The fake data points on the night most
affected by the planet perturbation lie significantly above the fitted point-lens
lightcurve, and these high points pull the fit up so that other points fall
systematically below the predicted lightcurve. As a result, the best fit achieved by the
point-lens no-planet model has a ${\chi^{2}/27=2.8}$ with 4 parameters fitted to 31 data
points. The 4 parameters were adjusted using the downhill simplex algorithm to 
minimize ${\chi^{2}}$ and were, namely, the time of maximum amplification, event
timescale, maximum amplification and baseline magnitude (${t_{0},t_{E},A_{0},I_{0}}$ 
respectively).

The ${\chi^{2}}$ improves by a factor of 8, to ${\chi^{2}/27=0.35}$ for a star+planet
lens model, as shown in Fig.~\ref{planet42}.
In this fit we adopt a planet/star mass ratio $q=10^{-3}$, and
allow the planet to be anywhere on the plane of the sky, thus
optimizing 2 additional parameters.
This highly significant improvement in the fit is sufficient to
reject the no-planet model in favor of the star+planet model.
This can also be seen clearly on the residual patterns for both fits as illustrated in 
figures \ref{resnoplan} and \ref{resplan} for the no planet and planet fit respectively.
The planet's presence is thus detectable in the lightcurve.

\begin{figure}
\centering
\begin{tabular}{c}
\psfig{file=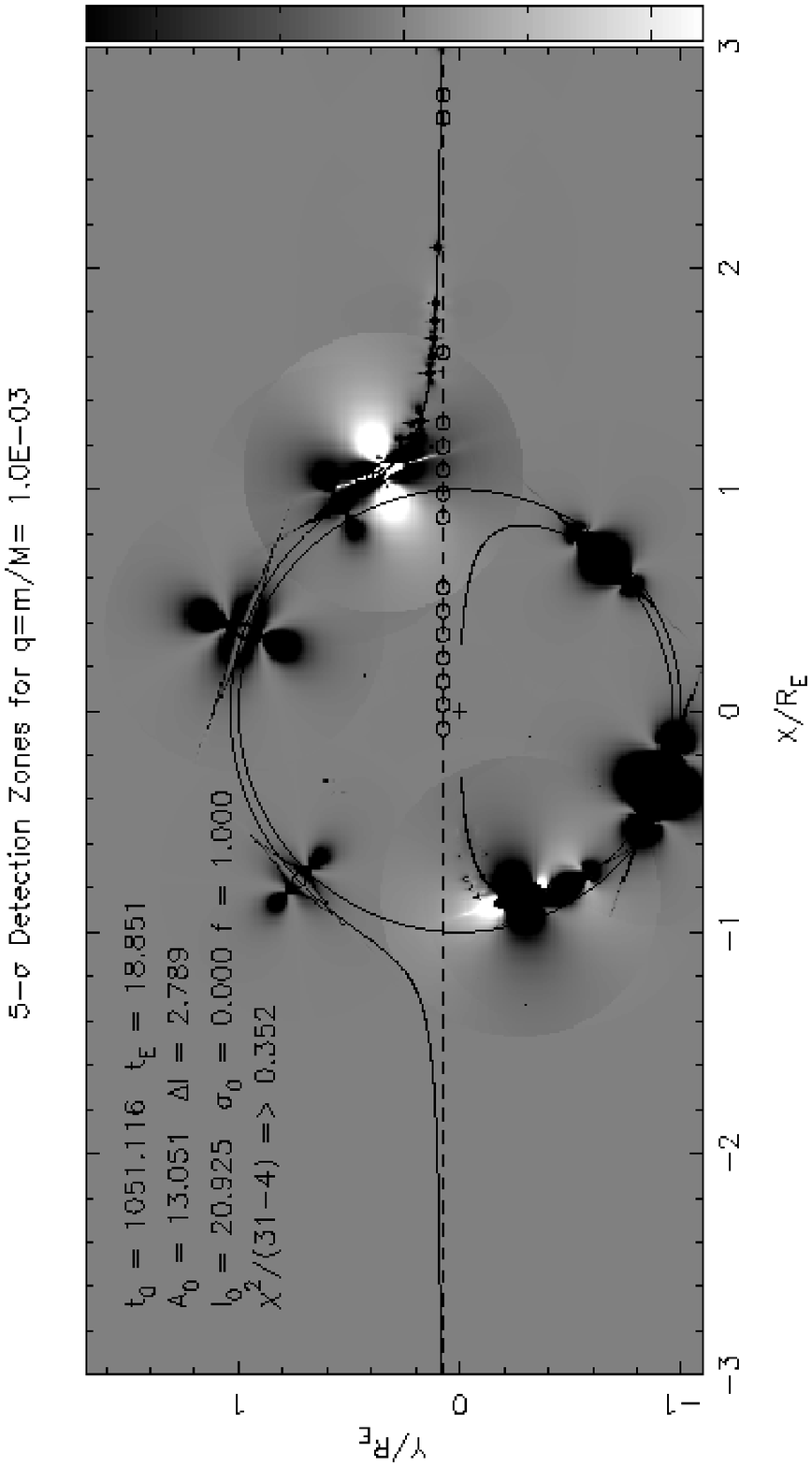,angle=-90,width=8cm}
\end{tabular}
\caption{\small The dark zones on the $\chi^2$ map mark where the planets with $q = 10^{-3}$ are 
excluded at the 5$\sigma$ level based
on the simulated observations. White zones represent a successful 
detection. The planet is successfully detected at position ${x/R_{E}=1.05}$, ${y/R_{E}=0.39}$
 on the lens plane where it interferes with one of the major images. Note that
 white detection zones also exist close to the minor image as well. This is
 because poor sampling cannot tell us exactly which image the planet is
 interacting with so we get white spots at both possible positions. This
 discrepancy can be solved if the sampling is sufficient to resolve the
 structure of the planet lensing event in detail.}
\protect\label{planetfound}
\end{figure}

Fig.~\ref{planetfound} shows the ${\Delta\chi^2}$ map as a function of assumed planet position.
Although the planet is detected, its mass and location are not well
defined from the data.
The data points that detect significant deviation from the point-lens lightcurve
do not reveal the duration or shape of the planetary deviation.
The planet could be interacting with either the major or minor images
of the source star, and therefore could be located on either of several positions
indicated by the white regions on Fig.~\ref{planetfound}.
Thus while the planet is detected, it is certainly not characterized.
Characterization obviously requires significantly more data points
to record the shape of the planetary deviation.

\begin{figure}
\centering
\begin{tabular}{c}
\psfig{file=planet42.ps,angle=-90,width=8cm}
\end{tabular}
\caption{\small Best fit lens+planet model for a simulated lightcurve which
includes a planetary deviation. This gives a lower ${\chi^{2}}$ than figure
\ref{single}, indicating a better fit. The fitted parameters appear in the top
left corner of the plot and the planet is at position ${x/R_{E}=1.05}$, ${y/R_{E}=0.39}$
 on the lens plane interfering with one of the major images.} 
\protect\label{planet42}
\centering
\begin{tabular}{c}
\psfig{file=res42pf.ps,angle=-90,width=8cm}
\end{tabular}
\caption{\small Fit residuals for the fit including the planet in fig~\ref{planet42}.}
\protect\label{resplan}
\end{figure} 

Since up to now there have been no confirmed reports 
of any planetary deviations by any microlensing follow-up network, it is our
belief that nightly monitoring schemes, taking a couple of exposures per night for a 
number  of events (as suggested in section 2) might yield the first detections. Even more so if numerous
telescopes contribute observations to the effort and data are shared in a common database.

\section{CONCLUSION}

We have used 1 hour per night on the IAC 0.8m telescope in Tenerife
for CCD monitoring of the lightcurves of Galactic Bulge microlensing
events during the 1998 season.
The best observed event in our dataset is 98BLG42, for which we
obtain accurate measurements on 4 consecutive nights
beginning just before the peak of the event,
and lower accuracy measurements in the tail of the event.
Our data are consistent with a point lens lightcurve.
We identify the detection zones near the Einstein ring of the lens star
where our data rule out the presence of a planet with a Jupiter-like
planet/star mass ratio $q=10^{-3}$.
For such planets our detection probability is 10\% for
orbit radius $a \sim R_E$, falling off for larger and smaller orbits.

We also demonstrate explicitly, by adding a few fake data points
to our actual CCD data, the feasibility of detecting
planets by monitoring microlensing lightcurves from small
(1m) telescopes at northern sites, despite the degradation of
accuracy arising from poorer seeing at higher airmass.

If such an observing scheme is to be pursued, ongoing events could be preselected from the 
alerts issued by the detection teams (OGLE, EROS) and observations could be directed to those of 
high amplification 
since the signal-to-noise (S/N) achieved should be better for those. Dense sampling should
be dedicated to clearly  defining the primary peak and probing for secondary peaks
in this region. If the lensing  star has a planetary companion, the probability of
detecting it is highest if the planet  has an orbital radius ${a \simeq R_{E}}$,
the Einstein ring radius. In this case the  planet could be perturbing either the
minor or the major image, which are located respectively just inside or just outside the 
Einstein ring at the high amplification part. Since the detection probability is much 
lower for ${a > > R_{E}}$ the event need not be monitored as densely for 
amplifications less than 1.34, where only a few data points are needed to
establish the  baseline level. The possibility of making observations from
northern sites may also yield  crucial data points on events that cannot be
followed during certain times from southern  sites where most teams currently
operate.

\section{ACKNOWLEDGEMENTS}

The data reductions were carried out at the St.Andrews node of the PPARC
Starlink Project. RAS was funded by a PPARC research studentship during the
course of this work. The IAC80 telescope is operated at Izana Observatory, Tenerife
by the Instituto de Astrofysica de Canarias.

\bibliographystyle{mn}
\bibliography{iau_journals,master,references}

\begin{thebibliography}{{Albrow {\rm et~al.}}{1998}}

\bibitem[\protect\citefmt{Alard \& Lupton}{1998}]{alard98}
Alard~C., Lupton~R., 1998, ApJ, 503, 325

\bibitem[\protect\citefmt{Albrow {\rm et~al.}}{1998}]{albrow98}
Albrow~M. {\rm et~al.}, 1998, ApJ, 509, 687

\bibitem[\protect\citefmt{Alcock~et al}{1997}]{MACHO}
Alcock~et al~C., 1997, ApJ, 479, 119

\bibitem[\protect\citefmt{Bennet \& Rhie}{1996}]{bennet96}
Bennet~D., Rhie~S., 1996, ApJ, 472, 660

\bibitem[\protect\citefmt{Gould \& Loeb}{1992}]{gould92}
Gould~A., Loeb~A., 1992, ApJ, 396, 104

\bibitem[\protect\citefmt{Marcy \& Butler}{1996}]{marcy96}
Marcy~G., Butler~R., 1996, ApJ, 464, 147

\bibitem[\protect\citefmt{Paczynski}{1986}]{pacz86}
Paczynski~B., 1986, ApJ, 304, 1

\bibitem[\protect\citefmt{Paczynski}{1996}]{pacz96}
Paczynski~B., 1996, ARA\&A, 34, 419

\bibitem[\protect\citefmt{Penny}{1995}]{starman}
Penny~A., 1995, Starlink User Note~141.2, Rutherford Appleton Laboratory

\bibitem[\protect\citefmt{Rhie \& Bennet}{1998}]{rhie98}
Rhie~S., Bennet~D., 1998, {\rm 193,}

\bibitem[\protect\citefmt{Udalski~et al}{1994}]{OGLE}
Udalski~et al~A., 1994, ApJ, 436, L103

\bibitem[\protect\citefmt{Wolzczan \& Frail}{1992}]{wolsz}
Wolzczan~A., Frail~D., 1992, Nat, 355, 145

\end{thebibliography}

\bsp
\end{document}